\documentclass[final]{IEEEtran}

\usepackage{amsmath, amssymb} 
\usepackage{graphicx}
\usepackage{url}
\usepackage{algpseudocode}
\usepackage{algorithm}
\usepackage{tikz}

\usepackage{float}
\newfloat{algorithm}{t}{lop}

\newcommand*{\MyPath}{./}

\title{FLiER:
  Practical Topology Update Detection Using Sparse
  PMUs\thanks{%
    The information, data, or work presented herein was
    funded in part by the Advanced Research Projects Agency-Energy
    (ARPA-E), U.S.  Department of Energy, under Award Number
    DE-AR0000230. The information, data, or work presented herein
    was funded in part by an agency of the United States
    Government. Neither the United States Government nor any agency
    thereof, nor any of their employees, makes any warranty, express
    or implied, or assumes any legal liability or responsibility for
    the accuracy, completeness, or usefulness of any information,
    apparatus, product, or process disclosed, or represents that its
    use would not infringe privately owned rights. Reference herein to
    any specific commercial product, process, or service by trade
    name, trademark, manufacturer, or otherwise does not necessarily
    constitute or imply its endorsement, recommendation, or favoring
    by the United States Government or any agency thereof. The views
    and opinions of authors expressed herein do not necessarily state
    or reflect those of the United States Government or any agency
    thereof.}}

\author{C. Ponce\thanks{This research was conducted with Government
    support under and awarded by DoD, Air Force Office of Scientific
    Research, National Defense Science and Engineering Graduate
    (NDSEG) Fellowship, 32 CFR 168a} \and D. S. Bindel}

\begin{document}

\maketitle

\begin{abstract}
In this paper, we present a Fingerprint Linear Estimation Routine (FLiER) to
identify topology changes in power networks using readings from
sparsely-deployed phasor measurement units (PMUs).  When a power line, load, or
generator trips in a network, or when a substation is reconfigured, the event
leaves a unique ``voltage fingerprint'' of bus voltage changes that we can
identify using only the portion of the network directly observed by the PMUs.
The naive brute-force approach to identify a failed line from such voltage
fingerprints, though simple and accurate, is slow.  We derive an approximate
algorithm based on a local linearization and a novel filtering approach that is
faster and only slightly less accurate.  We present experimental results using
the IEEE 57-bus, IEEE 118-bus, and Polish 1999-2000 winter peak networks.
\end{abstract}

\begin{IEEEkeywords}
  topology changes, phasor measurement units, voltage fingerprint,
  approximation, linearization, filtering
\end{IEEEkeywords}

\section{Introduction}

\subsection{Motivation}

Detection of topology changes is an important network monitoring function,
and is a key part of the power grid state estimation pipeline,
either as a pre-processing step or as an integrated part of a generalized state
estimator.  If a topology error processing module fails to detect an
error in the model topology,
poor and even dangerous control actions may
result~\cite{Ashok:2012:Cyber,Caro:2010:Breaker}, as unexpected topology
changes, such as those due to failed lines, may put stress on the remaining
lines and destabilize the network.  Thus, it is important to identify topology
changes quickly in order to take appropriate control actions.

Substations and transmission lines in transmission networks have sensors
that directly report failures (or switch
open/closed status). However, if a sensor malfunctions, then finding the
topology change is again difficult.  This can happen due to normal equipment
malfunctions, or because a cyber-attacker wishes to mislead network operators.
Although failure to correctly identify a topology error is less common in a
transmission network, the stakes are higher: state estimation based on incorrect
topology assumptions can lead to incorrect estimates, causing operators to
overlook system instability, and in the worst case, leading to avoidable
blackouts.  Thus, it is important to have more than one way to monitor network
topology.

\subsection{Prior work}

State estimation and topology detection have
co-evolved since at least the late 1970s~\cite{Lugtu:1980:Power};
see~\cite{Monticelli:2000:Electric} for an overview of the state
of the art as of 2000.  These industry-standard methods use
low time-resolution SCADA data about power flows and digital status,
together with a reference model, to infer system voltages and currents
as well as the current topology.
Since the turn of the century, researchers have increasingly proposed
PMU-based methods for state estimation, model calibration, fault detection,
and wide area monitoring and
control~\cite{7005374,5447627,Phadke:2008:Synchronized}.
Some PMU-based methods yield state
estimates~\cite{Yang:2011:TransitionI,Yang:2011:TransitionII},
information about oscillations~\cite{4610645,6072306,6246656,6299000},
and indicators of faults or contigencies~\cite{7169632,6808416}
without SCADA data and with little or no reference to a system model
during regular operation.  These ``model-free'' methods are attractive
because power system models are often not wholly correct.

Current PMU deployments do not provide complete observability of most of the
power grids, which has lead to a broad literature on optimally placing what
PMUs are available~\cite{5772601,6135526,6939388}.
And where PMU data is available, it is much
faster than other data sources: within an operating area,
SCADA sensors typically report at most every few seconds;
and the system data exchange (SDX) model of the North American
Electric Reliability Corporation (NERC) provides inter-area topology
information only on an hourly basis~\cite{nerc-report}.
According to the IEEE specification, data from a correctly functioning
PMU also satisfy tight phase and magnitude error tolerances~\cite{6839218}.
Hence, many methods combine PMU data, SCADA data, and
model information for state estimation~\cite{6807525,7112542,7166299,7273784,5608534,5871327,6384863,6407494,6465749,6570556,6878488,6953302,6845308} and for detection and
localization of faults~\cite{Tate:2008:Line,Tate:2009:Double,Zhu:2011:Lassoing,Zhu:2012:Sparse,7124531,7169632,7254202,6808416,6844899,6846291,6867388,6909088}.

Conventional state estimators use loss functions
such as least absolute variation (LAV) or Huber loss~\cite{Mili:1999:Robust,6807525}
for robustness to inconsistent outlier equations
caused by bad data or by errors in the model topology.
The shape of residuals associated with the outlier equations
reflects topology errors, and a quarter century of work
on topology estimators exploits this~\cite{Irving:1982:Substation,
Clements:1988:Detection,Wu:1989:Detection,Costa:1993:Identification,
Abur:1995:Identifying,Singh:1995:Network}.  Robust regression is
also used in many hybrid state
estimators~\cite{6807525,7112542,5608534,6878488},
as deployed PMUs sometimes have errors greater than those
nominally allowed by the IEEE
standard~\cite{1514485,4275758,6279481,6459588}, whether due
to poor GPS synchronization (which may cause persistent phase errors),
deficiencies in the signal processing algorithms (which may cause
oscillations in the PMU output), or other issues.  The methods in
this paper use differences between PMU predictions, so are insensitive
to absolute phase errors; and we filter the signal to reduce sensitivity
to deficiencies in signal processing algorithms and to ambient oscillations.
Nonetheless, we also use robust regression to defend against other types of
sensor errors.

Because PMUs measure current and voltage phasors directly,
PMU-assisted methods can observe topology changes
even in the sparse case when there is not a measurement directly
incident on the affected
element.  This is the basis for several
methods for line outage detection; the closest to our work
is~\cite{Tate:2008:Line}, though other closely related work
includes~\cite{Tate:2009:Double,Zhu:2011:Lassoing,Zhu:2012:Sparse}.
Our work extends these prior approaches by using PMU signals to
identify not only line trips, but also load trips, generator trips,
and substation reconfigurations.
As with many SCADA topology estimation techniques, we
model possible substation reconfigurations using a breaker-level
model~\cite{Monticelli:1991:Modeling,Monticelli:1993:Modeling,
Clements:1998:Topology,Mili:1999:Robust,Lourenco:2006:Topology}.
In the SCADA literature, one sometimes uses expanded bus-section
models only in regions where a substation reconfiguration is
suspected~\cite{Alsac:1998:Generalized}, but we use a fixed breaker-level
model and an extended system
with a minimal set of multipliers corresponding to the edges in trees
associated with each connected set of bus
sections~\cite{Gomez:2001:Reduced,DeLaVillaJaen:2002:Implicitly}.

\subsection{Our work}

We present here an efficient method to identify topology changes in
networks with a (possibly small) number of PMUs. We assume that a
complete state estimate is obtained shortly before a topology change,
e.g.~through conventional SCADA measurements, and we use discrepancies
between this state estimate and PMU measurements to identify failures.
Our method does not require complete observability from PMU data; it
performs well even when there are few PMUs in the network, though
having more PMUs does improve the accuracy.
Though our approach is similar in spirit to~\cite{Tate:2008:Line},
this paper makes three key novel contributions:
\begin{itemize}
\item We treat not only line outages, but also
  load trips, generator trips, and substation reconfigurations.
  This is not new for standard topology estimators,
  but to our knowledge it is new for PMU-based
  methods.
\item We describe a novel subspace-based filtering method to
  rule out candidate topology changes at low cost.
\item We take advantage of existing state estimation procedures
  by linearizing the full AC power flow about a previously estimated
  state.  This increases accuracy compared to the DC approximation
  used in prior work.
\end{itemize}

\section{Problem Formulation and Fingerprints}

Let $y_{ik} = g_{ik} + \jmath b_{ik}$ denote the elements of the
admittance matrix $Y \in \mathbb{C}^{n \times n}$ in a bus-branch
network model; let $P_\ell$ and $Q_\ell$ denote the real and reactive
power injections
at bus $\ell$;
and let $v_\ell = |v_\ell| \exp(\jmath \theta_\ell)$ denote the
voltage phasor at bus $\ell$.  These quantities are related by the
power flow equations
\begin{align}\label{eq:powerflow}
  H(v; Y) - s & = 0
\end{align}
where
\begin{align}
  \begin{bmatrix}
    H_\ell \\
    H_{n+\ell}
  \end{bmatrix} &=
  \sum_{h=1}^n |v_\ell||v_h|
  \begin{bmatrix}
    g_{\ell h} & b_{\ell h} \\
   -b_{\ell h} & g_{\ell h}
  \end{bmatrix}
  \begin{bmatrix}
    \cos(\theta_{\ell h}) \\
    \sin(\theta_{\ell h})
  \end{bmatrix},
\end{align}
with $\theta_{\ell h} = \theta_\ell - \theta_h$ and
\begin{align}
  s =
  \begin{bmatrix}
    P_1 & \cdots & P_n & Q_1 & \cdots & Q_n
  \end{bmatrix}^T.
\end{align}
We note that $H$ is quadratic in $v$, but linear $Y$.

In a breaker-level model, we use a similar system in which variables
are associated with bus sections, and $H$ represents the power flows
when all breakers are open.  We then write the power flow equations as
\begin{equation} \label{eq:powerflow-tie}
  \begin{split}
    H(v; Y) + C \lambda - s &= 0 \\
    C^T v &= b,
  \end{split}
\end{equation}
where the constraint equations $C^T v = b$ have the form
\[
  c_k^T v = (e_i-e_j)^T v = v_i-v_j = b_k = 0,
\]
i.e.~voltage variable $j$ for a ``slave'' bus section is constrained
to be the same as voltage variable $i$ for
a ``master'' bus section.
In addition,
we include constraints of the form
\[
  c_k^T v = e_i^T v = b_k
\]
to assign a voltage magnitude at a PV bus or the phase angle at a slack bus.
We could trivially eliminate these constraints,
but keep them explicit for notational convenience.

Our goal is to use the power flow equations to diagnose topology
changes such as single line failures, substation reconfigurations,
or load or generator trips.
We assume the network remains stable and the state shifts from one
quasi-steady state to another.
In practice, of course, there may be oscillations, whether due to ringdown
after a topology change or ambient forcing; hence we recommend applying
a low-pass filter to extract the mean behavior at each quasi-steady state.
Under a topology update, the voltage vector shifts
from $v$ to $\hat{v} = v + \Delta v$.  We assume
$m$ voltage phasor components, indicated by the rows of
$E \in \{0,1\}^{m   \times n}$, are directly observed by PMUs,
with appropriate filtering.
Assuming the loads and generation
vary slowly, modulo high-frequency fluctuations removed by the
low-pass filter, we can
predict what $E \Delta v$ should be for each possible contingency.
That is, we can match the observed voltage changes $E \Delta v$ to a
list of {\em voltage fingerprints}
to identify simple topology changes.  We note that the same approach
used to produce fingerprints for load and generator trips can also
be used to identify significant changes in load or generation at
a single source.

Multiple contingencies can have the same or
practically indistinguishable fingerprints.  For example, one of two
parallel lines with equal admittance may fail, or two lines that are
distant from all PMUs but near each other may yield similar
fingerprints. But even when a contingency is not identifiable, our
method still produces valuable information. When multiple lines have
the same effect on the network, our technique can be used to identify
a small set of potential lines or breakers to inspect more closely.

\section{Approximate Fingerprints} \label{sec:approx}

To compute the exact fingerprint for a contingency, we require a
nonlinear power flow solve.  In a large network with many possible
contingencies, this computation
becomes
expensive.  We approximate the
changing voltage in each contingency by linearizing the AC power flow
equations about the pre-contingency state.  As in methods based on the
DC approximation, we use the structure of changes to the
linearized system to compute voltage change fingerprints for each
contingency with a few linear solves.
By using
information
about the current
state,
we observe better diagnostic accuracy
with our AC linearization than with the DC approximation.

We consider three different types of contingencies: bus merging or bus
splitting due to substation reconfiguration, and line failure.  In
each case, we assume the pre-contingency state is $x = (v, \lambda)$
satisfying~\eqref{eq:powerflow-tie}.  We denote the post-contingency
state by primed variables $x' = (v', \lambda')$; we assume in general
that the power injections $s$ are the same before and after the
contingency.  The exact shift in state is $\Delta x' = x'-x$, and our
approximate fingerprints are based from the approximation
$\delta x' \approx \Delta x$ to the shift in state.  The computation of
$\delta x'$ for each contingency involves the pre-contingency
Jacobian matrix
\[
  A =
  \begin{bmatrix}
    \frac{\partial H}{\partial v}(v; Y) & C \\
    C^T & 0
  \end{bmatrix}.
\]
We assume a factorization of $A$ is available,
perhaps from a prior state estimate.

\subsection{Bus Merging Fingerprints}

In the case of two bus sections becoming electrically tied due to a
breaker closing, we augment $C$ by two additional
constraints $C'$ to tie together the voltage magnitudes and phase
angles of the previously-separate bus sections.  That is, the
post-contingency state satisfies the augmented system
\begin{equation} \label{eq:bus-merge-system}
  \begin{split}
    H(v'; Y) + C \lambda' + C' \gamma - s &= 0 \\
    C^T v' &= b \\
    C'^T v' &= 0.
  \end{split}
\end{equation}
We linearize~\eqref{eq:bus-merge-system} about the original state
$x$ (with $\gamma = 0$); because the first two equations are
satisfied at this state, we have the approximate system
\begin{equation} \label{eq:bus-merge-lin-system}
  \begin{bmatrix}
    A & U \\
    U^T & 0
  \end{bmatrix}
  \begin{bmatrix}
    \delta x' \\ \gamma
  \end{bmatrix} =
  -\begin{bmatrix}
    0 \\ 
    C'^T v
  \end{bmatrix}, \quad U = \begin{bmatrix} C' \\ 0 \end{bmatrix}
\end{equation}
We then solve the system by block elimination to obtain
\begin{align} 
  \gamma &= (U^T A^{-1} U)^{-1} (C'^T v)
    \label{eq:bus-split-linear-ge-a} \\
  \delta x' &= -A^{-1} U \gamma
    \label{eq:bus-split-linear-ge-b}
\end{align}
The formulas~\eqref{eq:bus-split-linear-ge-a}--\eqref{eq:bus-split-linear-ge-b}
only require two significant linear solves (to evaluate $A^{-1} U$),
some dot products, and a $2 \times 2$ solve.

\subsection{Bus Splitting Fingerprints}\label{subsec:bus_split}

When a bus splits after a breaker opens, the post-contingency state
satisfies the augmented system
\begin{equation} \label{eq:bus-split-system}
  \begin{split}
    H(v'; Y) + C \lambda' - s &= 0 \\
    C^T v' + F \gamma &= b \\
    F^T \lambda' &= 0.
  \end{split}
\end{equation}
The slack variables $\gamma$ let the voltage phasor for a
``breakaway'' group of previously-slaved sections differ
from a phasor at the former master section.
%
The two columns of $F \in \{0,1\}^{n \times 2}$ indicate
rows of $C^T$ that constrain the breakaway voltage magnitudes
and the phase angles, respectively.
The third equation says no power flows across the open breaker.

We linearize~\eqref{eq:bus-split-system} about the original
state $x$ (with $\gamma = 0$); because the first two equations
are satisfied at this state, we have the approximate system
\begin{equation} \label{eq:bus-split-lin-system}
  \begin{bmatrix}
    A & U \\
    U^T & 0
  \end{bmatrix}
  \begin{bmatrix}
    \delta x' \\ \gamma
  \end{bmatrix} =
  -\begin{bmatrix}
    0 \\ 
    F^T \lambda
  \end{bmatrix}, \quad U = \begin{bmatrix} 0 \\ F \end{bmatrix}.
\end{equation}
The bordered systems~\eqref{eq:bus-split-lin-system}
has the same form as~\eqref{eq:bus-merge-lin-system};
and, as before, block Gaussian elimination requires
only two solves with $A$, some dot products, and a $2 \times 2$ system
solve.

\subsection{Load/Generator Trip Fingerprints}

When a load or generator trips offline, that bus becomes a zero-injection PQ
nodes. In the case of a PQ load tripping offline, the network itself does not
change, so we do not need to augment the matrix $A$
as is done in Equations \eqref{eq:bus-merge-lin-system} and
\eqref{eq:bus-split-lin-system}, but we compute the approximate fingerprint with
just $A$. Rather, it is the power injection vector $s$ that changes.

In the case of a generator at a PV bus tripping offline, we need to convert that
bus into a PQ bus with zero power injection. We write the augmented system
\begin{equation} \label{eq:pv-trip-system}
  \begin{split}
    H(v'; Y) + C \lambda' - s' &= 0 \\
    C^T v' + f \gamma &= b \\
    f^T \lambda' &= 0.
  \end{split}
\end{equation}
The slack variable $\gamma$
lets the voltage magnitude of the bus of interest shift. The vector $f$
is an indicator vector such that the third equation constrains the reactive
power injection slack variable to zero. Note that $s$
has also been changed to $s'$
to represent the real power injection shifting to zero.

The resulting system is
\begin{equation} \label{eq:pv-trip-lin-system}
  \begin{bmatrix}
    A & u \\
    u^T & 0
  \end{bmatrix}
  \begin{bmatrix}
    \delta x' \\ \gamma
  \end{bmatrix} =
  -\begin{bmatrix}
    0 \\ 
    f^T \lambda
  \end{bmatrix}, \quad u = \begin{bmatrix} 0 \\ f \end{bmatrix}.
\end{equation}
The bordered system \eqref{eq:pv-trip-lin-system} has the same form as
\eqref{eq:bus-split-lin-system} and \eqref{eq:bus-merge-lin-system}. In this
case, only one solve with $A$ is required.

\subsection{Line Failure Fingerprints}

In principle, line failures can be handled in the same way as
substation reconfigurations that lead to bus splitting: 
explicitly represent two nodes on a line that are normally connected
(physically corresponding to two sides of a breaker) with a multiplier
that forces them to be equal, and compute the fingerprint by an
extended system that negates the effect of that multiplier.
In practice, we may prefer to avoid the extra variables in this
model.  The following formulation requires no explicit extra variables
in the base model, and can be used with either a breaker-level model
or a bus-branch model with no breakers (i.e.~$C$ an empty
matrix).

For line failures, the admittance changes to $Y' = Y + \Delta Y'$
where $\Delta Y'$ is a rank-one update.  The post-contingency
state satisfies the system
\[
\begin{split}
  H(v'; Y') + C \lambda' - s &= 0 \\
  C^T v' &= b,
\end{split}
\]
and linearization about $x$ gives
\begin{equation} \label{eq:line-fail-lin-system1}
\begin{bmatrix}
  \frac{\partial H}{\partial v}(v; Y') & C \\ C^T & 0
\end{bmatrix}
\begin{bmatrix} \delta v' \\ \delta \lambda' \end{bmatrix} =
-\begin{bmatrix} H(v; \Delta Y') \\ 0 \end{bmatrix}
\end{equation}
where $H(v; \Delta Y') = H(v; Y')-H(v; Y)$.  As we show momentarily,
\[
  \frac{\partial H}{\partial v}(v; Y')-
  \frac{\partial H}{\partial v}(v; Y) =
  \frac{\partial H}{\partial v}(v; \Delta Y') = U^0 (V^0)^T.
\]
where $U^0$ and $V^0$ each have three columns.
That is, the matrix in the system~\eqref{eq:line-fail-lin-system1} is
a rank-three update to $A$.  We can solve such a system by
the Sherman-Morrison-Woodbury update formula,
also widely known as the Inverse Matrix Modification Lemma~\cite{Alsac:1983:Sparsity,Hager:1989:Updating}.
We use the equivalent extended system
\begin{equation} \label{eq:line-fail-lin-system}
  \begin{bmatrix}
    A & U \\
    V^T & -I
  \end{bmatrix}
  \begin{bmatrix}
    \delta x' \\ \gamma
  \end{bmatrix} =
  -\begin{bmatrix}
    r \\ 0
  \end{bmatrix},
\end{equation}
where
\[
  U = \begin{bmatrix} U^0 \\ 0 \end{bmatrix}, \quad
  V = \begin{bmatrix} V^0 \\ 0 \end{bmatrix}, \quad
  r = \begin{bmatrix} H(v; \Delta Y') \\ 0 \end{bmatrix}.
\]
We again solve by block elimination:
\begin{align} \label{eq:line-ge-1}
  \gamma &= (I + V^T A^{-1} U)^{-1} (V^T r) \\
  \delta x' &= -A^{-1}(r + U \gamma). \label{eq:line-ge-2}
\end{align}
The work to
evaluate~\eqref{eq:line-ge-1}--\eqref{eq:line-ge-2} is
three linear solves (for $A^{-1} U$), some dot
products, and a small $3 \times 3$ solve.
We will show momentarily how
to avoid the solve involving $r$.

We now show that the Jacobian matrix changes by a rank-3 update.
For a failed line between nodes $i$ and $k$,
the vector $H(v, \Delta Y')$ has only four nonzero entries:
\begin{align*}
  \check P_i \equiv H_i ~~~&= \check P_{ik} + g_{ii}' |v_i|^2 \\
  \check Q_i \equiv H_{i+n} &= \check Q_{ik} - b_{ii}' |v_i|^2 \\
  \check P_k \equiv H_k ~~~&= \check P_{ki} + g_{kk}' |v_k|^2 \\
  \check Q_k \equiv H_{k+n} &= \check Q_{ki} - b_{kk}' |v_k|^2,
\end{align*}
where
\begin{equation*}
  \begin{bmatrix} \check P_{ik} \\ \check Q_{ik} \end{bmatrix} \equiv
  |v_i| |v_k| 
  \begin{bmatrix} 
    g_{ik}' & b_{ik}' \\
   -b_{ik}' & g_{ik}'
  \end{bmatrix} 
  \begin{bmatrix} \cos(\theta_{ik}) \\ \sin(\theta_{ik}) \end{bmatrix},
\end{equation*}
and $\check P_{ki}, \check Q_{ki}$ are defined similarly.
Let
\[
  D_{ik} \equiv 
  \frac{\partial (\check P_i, \check P_k, \check Q_i, \check Q_k)}
       {\partial (\theta_i, \theta_k, |v|_i, |v|_k)} \in \mathbb{R}^{4 \times 4};
\]
by the chain rule, we can write $D_{ik} = U_{ik} V_{ik}^T$ where
\begin{align*}
  U_{ik} &\equiv
  \frac{\partial (\check P_i, \check P_k, \check Q_i, \check Q_k)}
       {\partial (\theta_{ik}, \log |v|_i, \log |v|_k)} \in \mathbb{R}^{4 \times 3} \\
  V_{ik}^T &\equiv
  \frac{\partial (\theta_{ik}, \log |v|_i, \log |v|_k)}
       {\partial (\theta_i, \theta_k, |v|_i, |v|_k)} \in \mathbb{R}^{3 \times 4}.
\end{align*}
More concretely, we have
\begin{align*}
  & U_{ik} = \\
  & \begin{bmatrix}
    -\check Q_i - b_{ii}' |v_i|^2 &
     \check P_i + g_{ii}' |v_i|^2 &
     \check P_i - g_{ii}' |v_i|^2 \\
     \check Q_k + b_{kk}' |v_k|^2 &
     \check P_k - g_{kk}' |v_k|^2 &
     \check P_k + g_{kk}' |v_k|^2 \\
     \check P_i - g_{ii}' |v_i|^2 &
     \check Q_i - b_{ii}' |v_i|^2 &
     \Check Q_i + b_{ii}' |v_i|^2 \\
    -\check P_k + g_{kk}' |v_k|^2 &
     \check Q_k + b_{kk}' |v_k|^2 &
     \Check Q_k - b_{kk}' |v_k|^2
  \end{bmatrix} \\
  & V_{ik}^T =
  \begin{bmatrix}
    1 & -1 & 0 & 0 \\
    0 & 0 & |v_i|^{-1} & 0 \\
    0 & 0 & 0 & |v_k|^{-1}
  \end{bmatrix}.
\end{align*}
Because $H(v, \Delta Y')$ does not depend on any voltage phasors
other than those at nodes $i$ and $j$, we may write
\begin{equation}
  \frac{\partial H(v; \Delta Y')}{\partial v}
  = E_{ik} D_{ik} E_{ik}^T
  = U^0 (V^0)^T
\end{equation}
where
\begin{equation}
  E_{ik} = \begin{bmatrix} e_i & e_k \\ & & e_i & e_k
  \end{bmatrix} \in \mathbb{R}^{2n \times 4}.
\end{equation}
and
\begin{equation}
  U^0 = E_{ik} U_{ik}, \quad
  V^0 = E_{ik} V_{ik}.
\end{equation}
Moreover, we note that
\[
  H(v, \Delta Y') = E_{ik}
  \begin{bmatrix}
    \check P_i \\ \check P_k \\ \check Q_i \\ \check Q_k
  \end{bmatrix} =
  U^{0} z, \quad z = \begin{bmatrix} 0 \\ 1/2 \\ 1/2 \end{bmatrix},
\]
so that we may rewrite~\eqref{eq:line-ge-2} as
\begin{equation} \label{eq:line-ge-2b}
  \delta x' = -A^{-1} U (z + \gamma).
\end{equation}

\section{Filtering}\label{sec:filtering}

In
Section~\ref{sec:approx}
we discussed how to approximate voltage
shifts $\delta v'$ associated with several types of contingencies.
This approach to predicting voltage changes costs less
than a nonlinear power flow solve, but may still be costly for a large
network with many contingencies to check.  In the current section we
show how to rule out contingencies without any solves by computing
a cheap lower bound
on the discrepancy between the observed
voltage changes and the predicted voltage changes under the
contingencies.

For each contingency, we define the {\em fingerprint score}
\begin{equation}\label{eq:fingerprint}
  t = \|E \Delta v - E \delta v'\|
\end{equation}
where $\Delta v$ is the observed voltage shift and $\delta v'$ is
the voltage shift predicted for the contingency.  For the
contingencies we have described, $E \delta v'$ has the form
\begin{equation} \label{eq:E-dv-form}
  E \delta v' = \bar{E} A^{-1} U \gamma
\end{equation}
where $\bar{E} = \begin{bmatrix} E & 0 \end{bmatrix}$ simply ignores
the multiplier variables $\lambda$, and $\gamma$ is some short vector
of slack variables.  The expression $\bar{E} A^{-1}$
does not depend on the contingency,
and can be pre-computed at the cost
of $m$ linear solves (one per observed phasor component).  After this
computation, the
main cost in
evaluating~\eqref{eq:E-dv-form} is the computation of $\gamma$,
which involves a contingency-dependent linear system with $A$ as
an intermediate step.  However, we do not need $\gamma$
for
the {\em filter score}
\begin{equation} \label{eq:filter}
  \tau = \min_{\mu} \| E \Delta v - \bar{E} A^{-1} U \mu \| \leq t.
\end{equation}
In the
Euclidean norm, $\tau$ is simply
the size of the residual in a least squares fit of $E \Delta v$ to the
columns of $\bar{E} A^{-1} U$, which can be computed quickly due to
the sparsity of $U$.
If $U$ is the augmentation matrix associated with contingency $i$, 
we refer to $\bar{E} A^{-1} U$ as its \emph{filtering subspace}.

Filter score computations are cheap; and if the filter score $\tau_i$
for contingency $i$ exceeds the fingerprint score $t_k$ for
contingency $k$, then we know
\[
  t_k < \tau_i \leq t_i,
\]
without ever computing $t_i$.  Exploiting this fact leads to 
the FLiER method (Algorithm~\ref{alg:fast_approx})%
\footnote{Example Python code of this algorithm can be found at
\url{https://github.com/cponce512/FLiER_Test_Suite}}.

\begin{algorithm}
  \caption{FLiER}
  \label{alg:fast_approx}
  \begin{algorithmic}
    \State Compute and store $\bar{E} A^{-1}$.
    \State For each contingency $i$, compute $\tau_i$ via~\eqref{eq:filter}.
    \State Order the contingencies in ascending order by $\tau$.
    \For{$\ell = 2, 3, \ldots$}
      \State Compute fingerprint score $t_{\ell}$
      \State Break if {$t_{\ell} < \tau_{\ell+1}$}
    \EndFor
    \State Return contingencies with computed $t_{\ell}$
  \end{algorithmic}
\end{algorithm}

We note that PQ load trips require no filtering, as they involve no change to
the system matrix.


Filter score computations are embarrassingly parallel and can be
spread across processors.  Nonetheless, for huge networks with many
PMUs and many contingencies, the filter computations might be deemed
too expensive for very rapid diagnosis (e.g.~in less than a second).
However, the concept of a filter subspace can be adapted to these
cases.  First, one can define a {\em coarse} filtering subspaces that
is the sum of the filtering subspaces for a set of contingencies.  For
example, one might define a coarse filtering subspace associated with
all possible breaker reconfigurations inside a substation.  The coarse
subspace filter score provides a lower bound on the filter scores (and
hence the fingerprint scores) for all contingencies in the set.
Hence, it may not even be necessary to compute individual filter
scores for all contingencies considered.  Second, one can work with a
{\em projected} filtering subspace $W^T (EA^{-1} U)$ where $W$ is a
matrix with orthonormal columns.  The distance from a projected
measurement vector to the projected filtering subspace again gives a
lower bound on the full filter score.  In addition to reducing the
cost of filter score computations, projections can also be used to
eliminate faulty or missing PMU measurements from consideration.

\section{Experiments}

Our standard experimental setup is as follows. For each possible
topology change, we compute and pass to FLiER both the full
pre-contingency state and the subset of the post-contingency state
that would be observed by the PMUs.  We test both with no noise
and with independent random Gaussian noise with standard deviation
$1.7 \cdot 10^{-3}$ ($\approx 0.1$ degrees for phase angles) added to
both the initial state estimate and the PMU readings.
%
In~\cite{Tate:2008:Line}, $0.1$ degrees of Gaussian random noise
was applied to phase angles, then smoothed by passing a simulated
time-domain signal through a low pass filter; we apply the noise
without filtering, so the effect is more drastic.

One of the possibilities FLiER checks is that there has been no
change; in this case, we use the norm of the fingerprint as both the
fingerprint score and the filter score.  By including this possibility
among those checked, FLiER acts simultaneously as a method for
topology change detection and identification.

We run tests on the IEEE 57 bus and 118 bus networks,
with three different PMU arrangements on each:
\begin{itemize}
\item {\bf Single}: Only one PMU is placed in the network, at a low-degree node
  (bus 35 in the 57-bus network and 65 in the 118-bus network, providing a total
  of 3 and 5 bus voltage readings, respectively). This represents a
  near-worst-case deployment for our method.
\item {\bf Sparse}: A few PMUs are placed about the network (on buses 4, 13, and
  34 in the 57-bus network and on buses 5, 17, 37, 66, 80, and 100 in the
  118-bus network, providing a total of 15 and 40 bus voltage readings,
  respectively).  We consider this a realistic scenario in which
  sparsely-deployed PMUs do not offer full network observability.
\item {\bf All}: PMUs are placed on all buses.
  Any error is due purely to the linear approximation.
\end{itemize}
We did not test  changes that cause convergence failure in our
power flow solver. We assume such contingencies result in
collapse without some control action.

\subsection{Accuracy}

\subsubsection{Line Failures}

\begin{figure}
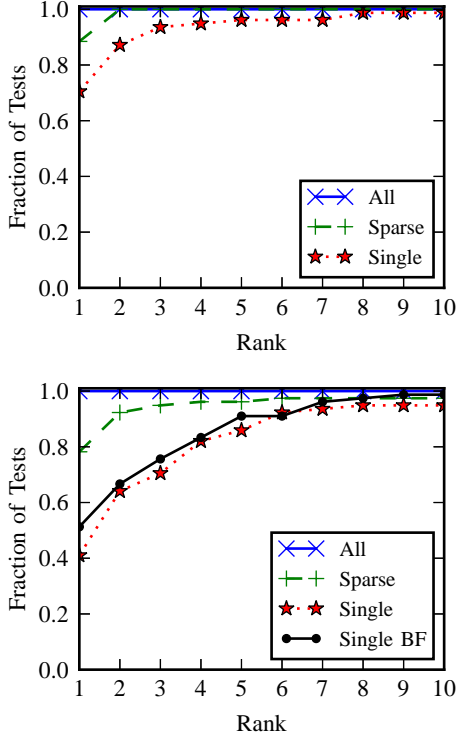

  \center
  \input{\MyPath data/ieee57ranks.pgf}
  \input{\MyPath data/ieee57ranks_noise0017.pgf}
  \caption{Cumulative distribution function showing the fraction of line
    failures where FLiER assigned the correct line at most a given rank (up to
    10).  Top: Noise-free case.  Bottom: Entries with Gaussian noise with
    $\sigma = 0.0017$.
    ``Single BF'' is the result from using a brute-force approach with a single
    PMU and represents the signal-to-noise ratio in that PMU's information.}
  \label{fig:ieee57ranks}
\end{figure}
Figure \ref{fig:ieee57ranks} shows the accuracy of FLiER in
identifying line failures in the IEEE 57-bus test network.  For each
PMU deployment, we show the cumulative distribution function of ranks,
i.e.~the ranks of each simulated contingency in the ordered list
produced by FLiER.  We show further results in
Table~\ref{tab:comparison}. With PMUs everywhere, the correct answer
was chosen in
all 78 of 78 cases, even in the presence of noise. The case with three
PMUs is also quite robust to noise.
In the test with a single unfavorably-placed PMU, FLiER typically
ranks the correct line among the top three in the absence of noise;
with noise, the accuracy degrades, though not completely.

\begin{figure}
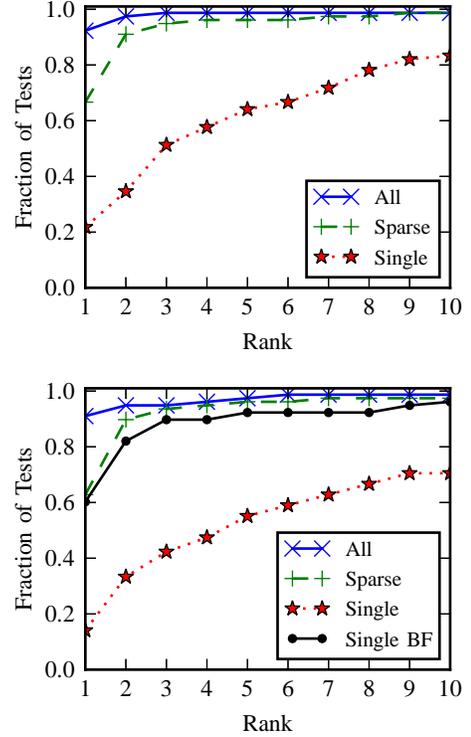

  \center
  \input{\MyPath data/ieee57ranksDC.pgf}
  \input{\MyPath data/ieee57ranksDC_noise0017.pgf}
  \caption{CDF of line failures where the DC approximation
    of~\cite{Tate:2008:Line} assigned the correct line at most a given rank (up
    to 10).  Top: Noise-free case.  Bottom: Entries with Gaussian noise with
    $\sigma = 0.0017$.
    ``Single BF'' is the result from using a brute-force approach with a single
    PMU and represents the signal-to-noise ratio in that PMU's information.}
  \label{fig:ieee57ranksDC}
\end{figure}

\begin{table}
  \centering
  \begin{tabular}{| l | r | r | r |}
    \hline
    PMUs & Single & Sparse & All \\
    \hline
    FLiER & 55(73) & 68(77) & 78(78) \\
    FLiER+noise & 40(65) & 66(78) & 78(78) \\
    DC Approx & 17(40) & 52(74) & 72(77) \\
    DC Approx+noise & 5(22) & 49(64) & 66(68) \\
    \hline
  \end{tabular}
  \caption{IEEE 57-bus network accuracy comparison for 78 line failure
    contingencies.  We report counts of line failures correctly
    identified and those scored in the top three (in parentheses).}
  \label{tab:comparison}
\end{table}

In Figure~\ref{fig:ieee57ranksDC}, we repeat the experiment of
Figure~\ref{fig:ieee57ranks}, but with the DC approximation used
in~\cite{Tate:2008:Line} rather than the AC linearization used in
FLiER.  We also present comparisons in
Table~\ref{tab:comparison}.  With PMUs everywhere, there is little
difference in accuracy.  With fewer PMUs, FLiER is more accurate.  In
the sparse case, the DC approximation without noise behaves similarly
to FLiER with noise, while in the single PMU deployment the DC results
without noise are much worse than those from FLiER even with noise.

\begin{figure}[t]
  \centering
  \input{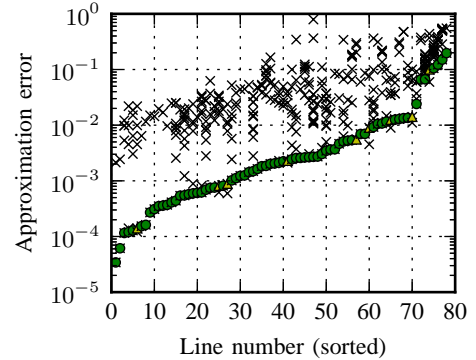}
  \caption{Test of our algorithm on the IEEE 57-bus network with the
    sparse PMU deployment.  Each column represents one test.
    Black crosses are fingerprint scores for incorrect lines.
    Green dots and yellow triangles indicate the scores of the correct
    line in the case of correct diagnosis or diagnosis in the top
    three, respectively.}
  \label{fig:pmus_31233}
\end{figure}

Figure \ref{fig:pmus_31233} shows the raw scores computed by FLiER
with three PMUs.  In this plot, each column represents the fingerprint
scores computed for one line failure scenario.  The black
crosses
represent the scores of lines that get past the filter, while
the green circles and yellow triangles represent the scores for the
correct answer.  If there is a green circle, then our algorithm
correctly identified the actual line that failed. If there is a yellow
triangle, the correct line was not chosen but was among the top three
lines selected by the algorithm.

\begin{figure}[t]
  \centering
  \scalebox{0.8}{\begin{tikzpicture}[xscale=0.075, yscale=0.06]
\input{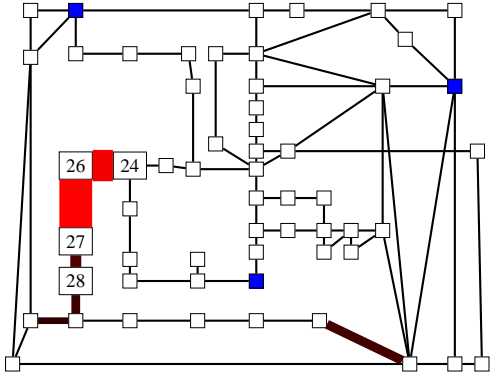}
\definecolor{currentcolor}{rgb}{0.0, 0.0, 0.0} ;
\draw[color=currentcolor, line width=1pt] (n0) -- (n16) ;
\definecolor{currentcolor}{rgb}{0.0, 0.0, 0.0} ;
\draw[color=currentcolor, line width=1pt] (n0) -- (n1) ;
\definecolor{currentcolor}{rgb}{0.0, 0.0, 0.0} ;
\draw[color=currentcolor, line width=1pt] (n0) -- (n14) ;
\definecolor{currentcolor}{rgb}{0.0, 0.0, 0.0} ;
\draw[color=currentcolor, line width=1pt] (n0) -- (n15) ;
\definecolor{currentcolor}{rgb}{0.0, 0.0, 0.0} ;
\draw[color=currentcolor, line width=1pt] (n1) -- (n2) ;
\definecolor{currentcolor}{rgb}{0.0, 0.0, 0.0} ;
\draw[color=currentcolor, line width=1pt] (n2) -- (n3) ;
\definecolor{currentcolor}{rgb}{0.0, 0.0, 0.0} ;
\draw[color=currentcolor, line width=1pt] (n2) -- (n14) ;
\definecolor{currentcolor}{rgb}{0.0, 0.0, 0.0} ;
\draw[color=currentcolor, line width=1pt] (n3) -- (n17) ;
\definecolor{currentcolor}{rgb}{0.0, 0.0, 0.0} ;
\draw[color=currentcolor, line width=1pt] (n3) -- (n4) ;
\definecolor{currentcolor}{rgb}{0.0, 0.0, 0.0} ;
\draw[color=currentcolor, line width=1pt] (n3) -- (n5) ;
\definecolor{currentcolor}{rgb}{0.0, 0.0, 0.0} ;
\draw[color=currentcolor, line width=1pt] (n4) -- (n5) ;
\definecolor{currentcolor}{rgb}{0.0, 0.0, 0.0} ;
\draw[color=currentcolor, line width=1pt] (n5) -- (n6) ;
\definecolor{currentcolor}{rgb}{0.0, 0.0, 0.0} ;
\draw[color=currentcolor, line width=1pt] (n5) -- (n7) ;
\definecolor{currentcolor}{rgb}{0.1771229, 0.0, 0.0} ;
\draw[color=currentcolor, line width=3.15684465662pt] (n6) -- (n28) ;
\definecolor{currentcolor}{rgb}{0.0, 0.0, 0.0} ;
\draw[color=currentcolor, line width=1pt] (n6) -- (n7) ;
\definecolor{currentcolor}{rgb}{0.0, 0.0, 0.0} ;
\draw[color=currentcolor, line width=1pt] (n7) -- (n8) ;
\definecolor{currentcolor}{rgb}{0.0, 0.0, 0.0} ;
\draw[color=currentcolor, line width=1pt] (n8) -- (n9) ;
\definecolor{currentcolor}{rgb}{0.0, 0.0, 0.0} ;
\draw[color=currentcolor, line width=1pt] (n8) -- (n10) ;
\definecolor{currentcolor}{rgb}{0.0, 0.0, 0.0} ;
\draw[color=currentcolor, line width=1pt] (n8) -- (n11) ;
\definecolor{currentcolor}{rgb}{0.0, 0.0, 0.0} ;
\draw[color=currentcolor, line width=1pt] (n8) -- (n12) ;
\definecolor{currentcolor}{rgb}{0.2581988, 0.0, 0.0} ;
\draw[color=currentcolor, line width=4.372983346205pt] (n8) -- (n54) ;
\definecolor{currentcolor}{rgb}{0.0, 0.0, 0.0} ;
\draw[color=currentcolor, line width=1pt] (n9) -- (n50) ;
\definecolor{currentcolor}{rgb}{0.0, 0.0, 0.0} ;
\draw[color=currentcolor, line width=1pt] (n9) -- (n11) ;
\definecolor{currentcolor}{rgb}{0.0, 0.0, 0.0} ;
\draw[color=currentcolor, line width=1pt] (n10) -- (n40) ;
\definecolor{currentcolor}{rgb}{0.0, 0.0, 0.0} ;
\draw[color=currentcolor, line width=1pt] (n10) -- (n42) ;
\definecolor{currentcolor}{rgb}{0.0, 0.0, 0.0} ;
\draw[color=currentcolor, line width=1pt] (n10) -- (n12) ;
\definecolor{currentcolor}{rgb}{0.0, 0.0, 0.0} ;
\draw[color=currentcolor, line width=1pt] (n11) -- (n16) ;
\definecolor{currentcolor}{rgb}{0.0, 0.0, 0.0} ;
\draw[color=currentcolor, line width=1pt] (n11) -- (n12) ;
\definecolor{currentcolor}{rgb}{0.0, 0.0, 0.0} ;
\draw[color=currentcolor, line width=1pt] (n11) -- (n15) ;
\definecolor{currentcolor}{rgb}{0.0, 0.0, 0.0} ;
\draw[color=currentcolor, line width=1pt] (n12) -- (n48) ;
\definecolor{currentcolor}{rgb}{0.0, 0.0, 0.0} ;
\draw[color=currentcolor, line width=1pt] (n12) -- (n13) ;
\definecolor{currentcolor}{rgb}{0.0, 0.0, 0.0} ;
\draw[color=currentcolor, line width=1pt] (n12) -- (n14) ;
\definecolor{currentcolor}{rgb}{0.0, 0.0, 0.0} ;
\draw[color=currentcolor, line width=1pt] (n13) -- (n45) ;
\definecolor{currentcolor}{rgb}{0.0, 0.0, 0.0} ;
\draw[color=currentcolor, line width=1pt] (n13) -- (n14) ;
\definecolor{currentcolor}{rgb}{0.0, 0.0, 0.0} ;
\draw[color=currentcolor, line width=1pt] (n14) -- (n44) ;
\definecolor{currentcolor}{rgb}{0.0, 0.0, 0.0} ;
\draw[color=currentcolor, line width=1pt] (n17) -- (n18) ;
\definecolor{currentcolor}{rgb}{0.0, 0.0, 0.0} ;
\draw[color=currentcolor, line width=1pt] (n18) -- (n19) ;
\definecolor{currentcolor}{rgb}{0.0, 0.0, 0.0} ;
\draw[color=currentcolor, line width=1pt] (n20) -- (n19) ;
\definecolor{currentcolor}{rgb}{0.0, 0.0, 0.0} ;
\draw[color=currentcolor, line width=1pt] (n20) -- (n21) ;
\definecolor{currentcolor}{rgb}{0.0, 0.0, 0.0} ;
\draw[color=currentcolor, line width=1pt] (n21) -- (n37) ;
\definecolor{currentcolor}{rgb}{0.0, 0.0, 0.0} ;
\draw[color=currentcolor, line width=1pt] (n21) -- (n22) ;
\definecolor{currentcolor}{rgb}{0.0, 0.0, 0.0} ;
\draw[color=currentcolor, line width=1pt] (n22) -- (n23) ;
\definecolor{currentcolor}{rgb}{0.0, 0.0, 0.0} ;
\draw[color=currentcolor, line width=1pt] (n23) -- (n24) ;
\definecolor{currentcolor}{rgb}{0.9558898, 0.0, 0.0} ;
\draw[color=currentcolor, line width=14.83834841585pt] (n23) -- (n25) ;
\definecolor{currentcolor}{rgb}{0.0, 0.0, 0.0} ;
\draw[color=currentcolor, line width=1pt] (n24) -- (n29) ;
\definecolor{currentcolor}{rgb}{1.0, 0.0, 0.0} ;
\draw[color=currentcolor, line width=15.5pt] (n25) -- (n26) ;
\definecolor{currentcolor}{rgb}{0.3131121, 0.0, 0.0} ;
\draw[color=currentcolor, line width=5.19668218315pt] (n26) -- (n27) ;
\definecolor{currentcolor}{rgb}{0.2425356, 0.0, 0.0} ;
\draw[color=currentcolor, line width=4.138034375545pt] (n27) -- (n28) ;
\definecolor{currentcolor}{rgb}{0.0, 0.0, 0.0} ;
\draw[color=currentcolor, line width=1pt] (n28) -- (n51) ;
\definecolor{currentcolor}{rgb}{0.0, 0.0, 0.0} ;
\draw[color=currentcolor, line width=1pt] (n29) -- (n30) ;
\definecolor{currentcolor}{rgb}{0.0, 0.0, 0.0} ;
\draw[color=currentcolor, line width=1pt] (n30) -- (n31) ;
\definecolor{currentcolor}{rgb}{0.0, 0.0, 0.0} ;
\draw[color=currentcolor, line width=1pt] (n31) -- (n32) ;
\definecolor{currentcolor}{rgb}{0.0, 0.0, 0.0} ;
\draw[color=currentcolor, line width=1pt] (n33) -- (n34) ;
\definecolor{currentcolor}{rgb}{0.0, 0.0, 0.0} ;
\draw[color=currentcolor, line width=1pt] (n33) -- (n31) ;
\definecolor{currentcolor}{rgb}{0.0, 0.0, 0.0} ;
\draw[color=currentcolor, line width=1pt] (n34) -- (n35) ;
\definecolor{currentcolor}{rgb}{0.0, 0.0, 0.0} ;
\draw[color=currentcolor, line width=1pt] (n35) -- (n36) ;
\definecolor{currentcolor}{rgb}{0.0, 0.0, 0.0} ;
\draw[color=currentcolor, line width=1pt] (n35) -- (n39) ;
\definecolor{currentcolor}{rgb}{0.0, 0.0, 0.0} ;
\draw[color=currentcolor, line width=1pt] (n36) -- (n37) ;
\definecolor{currentcolor}{rgb}{0.0, 0.0, 0.0} ;
\draw[color=currentcolor, line width=1pt] (n36) -- (n38) ;
\definecolor{currentcolor}{rgb}{0.0, 0.0, 0.0} ;
\draw[color=currentcolor, line width=1pt] (n37) -- (n48) ;
\definecolor{currentcolor}{rgb}{0.0, 0.0, 0.0} ;
\draw[color=currentcolor, line width=1pt] (n37) -- (n43) ;
\definecolor{currentcolor}{rgb}{0.0, 0.0, 0.0} ;
\draw[color=currentcolor, line width=1pt] (n37) -- (n47) ;
\definecolor{currentcolor}{rgb}{0.0, 0.0, 0.0} ;
\draw[color=currentcolor, line width=1pt] (n38) -- (n56) ;
\definecolor{currentcolor}{rgb}{0.0, 0.0, 0.0} ;
\draw[color=currentcolor, line width=1pt] (n39) -- (n55) ;
\definecolor{currentcolor}{rgb}{0.0, 0.0, 0.0} ;
\draw[color=currentcolor, line width=1pt] (n40) -- (n41) ;
\definecolor{currentcolor}{rgb}{0.0, 0.0, 0.0} ;
\draw[color=currentcolor, line width=1pt] (n40) -- (n42) ;
\definecolor{currentcolor}{rgb}{0.0, 0.0, 0.0} ;
\draw[color=currentcolor, line width=1pt] (n43) -- (n44) ;
\definecolor{currentcolor}{rgb}{0.0, 0.0, 0.0} ;
\draw[color=currentcolor, line width=1pt] (n45) -- (n46) ;
\definecolor{currentcolor}{rgb}{0.0, 0.0, 0.0} ;
\draw[color=currentcolor, line width=1pt] (n46) -- (n47) ;
\definecolor{currentcolor}{rgb}{0.0, 0.0, 0.0} ;
\draw[color=currentcolor, line width=1pt] (n47) -- (n48) ;
\definecolor{currentcolor}{rgb}{0.0, 0.0, 0.0} ;
\draw[color=currentcolor, line width=1pt] (n48) -- (n49) ;
\definecolor{currentcolor}{rgb}{0.0, 0.0, 0.0} ;
\draw[color=currentcolor, line width=1pt] (n49) -- (n50) ;
\definecolor{currentcolor}{rgb}{0.0, 0.0, 0.0} ;
\draw[color=currentcolor, line width=1pt] (n51) -- (n52) ;
\definecolor{currentcolor}{rgb}{0.0, 0.0, 0.0} ;
\draw[color=currentcolor, line width=1pt] (n52) -- (n53) ;
\definecolor{currentcolor}{rgb}{0.0, 0.0, 0.0} ;
\draw[color=currentcolor, line width=1pt] (n53) -- (n54) ;
\definecolor{currentcolor}{rgb}{0.0, 0.0, 0.0} ;
\draw[color=currentcolor, line width=1pt] (n55) -- (n40) ;
\definecolor{currentcolor}{rgb}{0.0, 0.0, 0.0} ;
\draw[color=currentcolor, line width=1pt] (n55) -- (n41) ;
\definecolor{currentcolor}{rgb}{0.0, 0.0, 0.0} ;
\draw[color=currentcolor, line width=1pt] (n56) -- (n55) ;
\end{tikzpicture}}
  \caption{Line (24, 26) is the line removed in this test. Lines are
    colored and thickened according to $\sqrt{t_{ik}^{-1}}$. Line
    (26,~27) was chosen by the algorithm.}
  \label{fig:24_26_down}
\end{figure}

In Figure \ref{fig:24_26_down}, we show one case that FLiER
misidentifies.
PMUs are deployed on buses marked with blue squares,
and lines are colored and thickened according to the FLiER score.
The best-scoring line is adjacent to the line that failed.

\subsubsection{Substation Reconfigurations}

Next, we show the accuracy of FLiER as it applies to substation
reconfigurations. For these tests, we suppose that every bus in the
IEEE 57-bus test network is a ring substation with each bus section on
the ring possessing either load, generation, or a branch. We then
suppose a substation splits when two of its circuit breakers open. We
do not consider cases that isolate a node with a nonzero power
injection.  Line failures are a subset of this scenario: if
the breakers on either side of a section with a branch open, that
section becomes a zero-injection leaf bus, which disappears in the
quasi-static setting.

\begin{figure}[t]
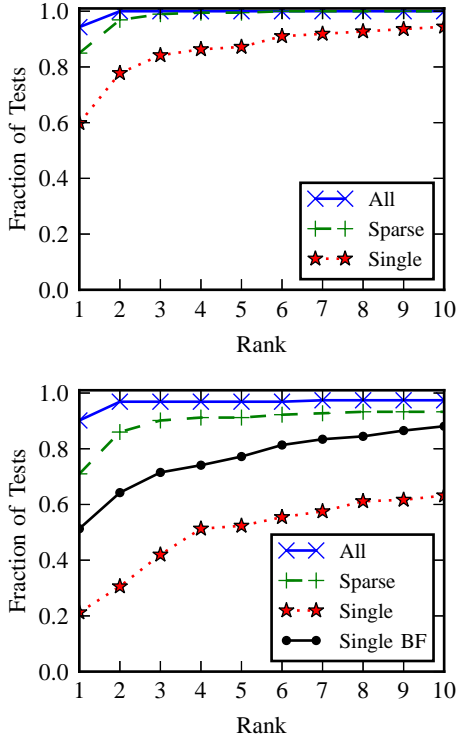

  \center
  \input{\MyPath data/ieee57subranks.pgf}
  \input{\MyPath data/ieee57subranks_noise0017.pgf}
  \caption{Rank CDF for substation reconfigurations without noise (top) and with
    noise (bottom).  ``Single BF'' is the result from using a brute-force
    approach with a single PMU and represents the signal-to-noise ratio in that
    PMU's information.}
  \label{fig:ieee57subranks}
\end{figure}

Figure \ref{fig:ieee57subranks} shows the accuracy of FLiER on
substation reconfigurations with and without noise. With three PMUs
and no noise, FLiER
is right
in 164 of 193
possibilities, and ranks the correct answer among the top three scores
in 160 cases. With PMUs everywhere, FLiER
is right
181 times, but gets the answer in the top three every single time.
With few PMUs, FLiER is more susceptible to noise when diagnosing
substation reconfigurations. This is expected, as there are
significantly more possibilities to choose from in this case.
Also, FLiER
sometimes
filters out the correct answer in the
presence of noise. One possible remedy for this would be to be more
lenient with filtering, only throwing a possibility away if
$\tau_\ell$ is greater than the $k$th smallest $t_\ell$, for example.

Finally, we demonstrate the effectiveness of using FLiER for substation
reconfigurations on a large-scale network
by running FLiER on the 400, 220, and 100 kV subset of the Polish
network during peak conditions of the 1999-2000 winter, taken
from~\cite{Zimmerman:2011:MATPOWER}. This is a larger network with
2,383 buses. We placed 100 PMUs randomly
around the network, and tested every substation reconfiguration
contingency.
We summarize the results
in Table
\ref{tab:polish_accuracy}.
We could likely further improve the accuracy with a thoughtful
deployment of PMUs.

\begin{table}
   \centering
   \begin{tabular}{| l | r | r |}
\hline
            & Contingency & Substation \\
           &                     & of contingency \\
\hline
Correct \%  & 75.2   &   85.4  \\
Top 3 \%     & 95.4   &   96.5  \\
\hline
   \end{tabular}
   \caption{Accuracy of FLiER with 100 randomly-placed PMUs on the
               Polish network. Results are out of 6283 tests.}
   \label{tab:polish_accuracy}
\end{table}

\subsubsection{Including Load and Generator Trips}

\begin{figure}[t]
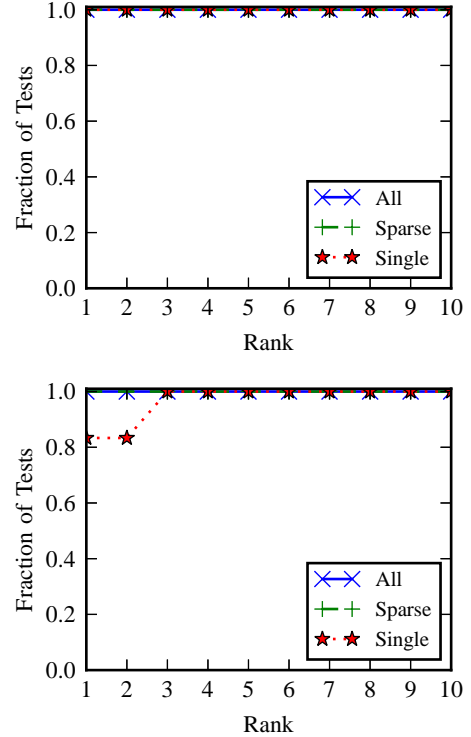

  \center
  \input{\MyPath data/ieee57gendropranks.pgf}
  \input{\MyPath data/ieee57gendropranks_noise0017.pgf}
  \caption{Rank CDF for generator trips without noise (top) and with noise
    (bottom).}
  \label{fig:ieee57gendropranks}
\end{figure}

\begin{figure}[t]
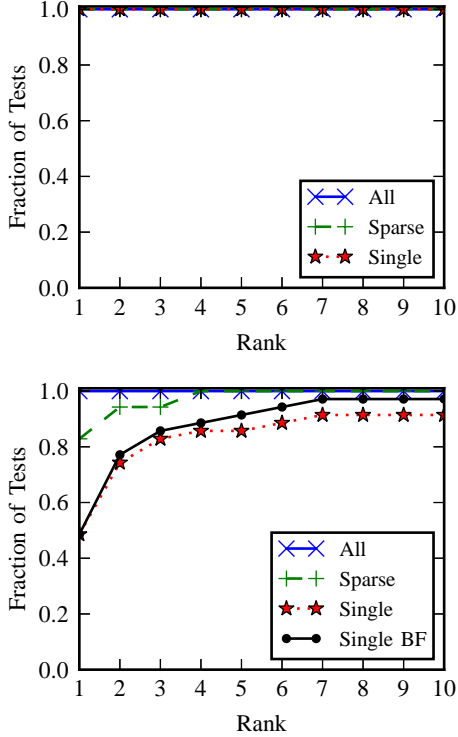

  \center
  \input{\MyPath data/ieee57loaddropranks.pgf}
  \input{\MyPath data/ieee57loaddropranks_noise0017.pgf}
  \caption{Rank CDF for load trips without noise (top) and with noise
    (bottom). In the noisy case, we allow a filtering slack equal to one noise
    standard deviation, as otherwise the filter is too stringent. ``Single BF''
    is the result from using a brute-force approach with a single PMU and
    represents the signal-to-noise ratio in that PMU's information. We do not
    include zero-injection busses in this test.}
  \label{fig:ieee57loaddropranks}
\end{figure}

\begin{figure}[t]
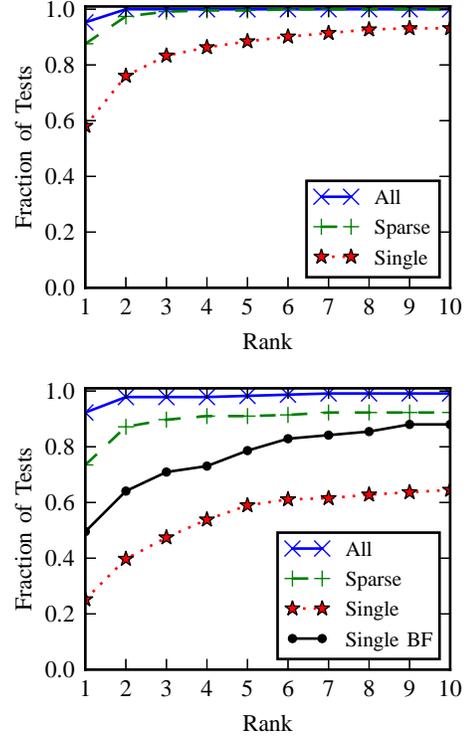

  \center
  \input{\MyPath data/ieee57allranks.pgf}
  \input{\MyPath data/ieee57allranks_noise0017.pgf}
  \caption{Rank CDF for substation reconfigurations and load/generator trips
    without noise (top) and with noise (bottom). ``Single BF'' is the result
    from using a brute-force approach with a single PMU and represents the
    signal-to-noise ratio in that PMU's information.}
  \label{fig:ieee57allranks}
\end{figure}

Here we include load and generator trips in our results. In Figure
\ref{fig:ieee57gendropranks}, we show the results of testing for generator
trips, in Figure \ref{fig:ieee57loaddropranks}, we show the results of testing
for load trips, and in Figure \ref{fig:ieee57allranks}, we show the results of
including load and generator trips in the set of contingencies to test and check
for along with substation splits. All tests use the IEEE 57-bus network.

In the noisy case for Figure \ref{fig:ieeeloaddropranks}, we allow some extra
slack in the filtering procedure. In particular, rather than checking if
$t_k < \tau_i$,
we check if $t_k < \tau_i - \sigma$,
where $\sigma$
is the standard deviation of the included noise. We did this because the
filtering method sometimes filtered out the correct answer for load trips. In a
real-world setting one would need to choose that slack value intelligently, but
this test shows that slightly less-stringent filtering can improve the method in
some cases.

\subsubsection{PMU Placement}

\begin{figure}[t]
  \center
  \input{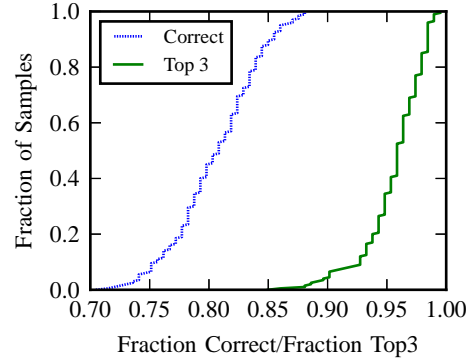}
  \caption{CDF for fraction of substation reconfiguration contingencies that
    FLiER gets correct (blue) and in the top 3 (green) with random uniform
    placement of three PMUs. Noise is not included in this test.}
  \label{fig:ieee57random_pmus}
\end{figure}

Here we demonstrate the robustness of FLiER to PMU placement. We tested this by
sampling three busses from the IEEE 57-bus network uniformly at random, running
FLiER for the substation reconfiguration case (as in Figure
\ref{fig:ieee57subranks}) and recording the fraction of contingencies FLiER
identified correctly, and the fraction of contingencies with the solution ranked
in the top 3. We performed this test 200 times. We show cumulative distribution
functions for the fraction correct and fraction in the top 3 in Figure
\ref{fig:ieee57random_pmus}.

This figure shows that FLiER tends to be quite robust to PMU placement. In fact,
the median fraction correct and median fraction in top 3 are only slightly below
those for the noiseless test in Figure \ref{fig:ieee57subranks}, where we
attempted to place the three PMUs favorably. Furthermore, the standard deviation
for fraction correct is only 3.65\%, while the standard deviation for fraction
in top 3 is 2.67\%.

\subsubsection{Robustness to Bad Data}

Here we demonstrate how one can easily modify FLiER for enhanced robustness to
bad data and heavy-tailed noise. In Equations \eqref{eq:fingerprint} and
\eqref{eq:filter}, we use the L2 loss function to measure the distance between
an estimated fingerprint or subspace and the actual fingerprint. The L2 loss
function, while simple and useful, is sensitive to outliers. A single large
error component drastically increases the $t$
or $\tau$
value, even if all other components have very small error. The result is using
the L2 loss can make FLiER sensitive to bad data and heavy-tailed noise.

An alternative is to use a loss function that is robust to large outliers. One
popular such loss function is the \emph{Huber loss}~\cite{Mili:1999:Robust}:
\begin{align}
\| e \|_{H}^2 = & \sum_{i=1}^m L_\delta(e_i) \\
L_\delta(e_i) = & \left\{ \begin{array}{ll}
\frac{1}{2} e_i^2 & |e_i| \leq \delta \\
\delta \left(|e_i| - \frac{1}{2} \delta \right) & \mbox{otherwise}
\end{array} \right.
\end{align}
This function behaves like the L2 loss for error components near zero, but for
error components larger than $\delta$,
the loss grows linearly rather than quadratically. The result is a loss function
that is robust to outliers in data.

\begin{figure}[t]
   \center
   \input{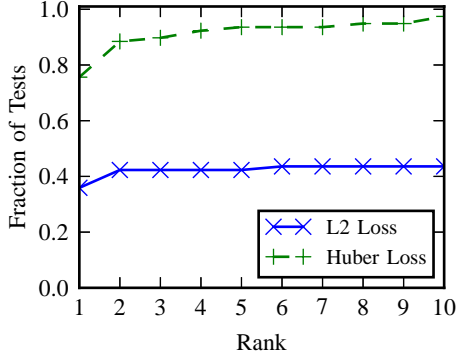}
   \caption{CDF for line failures with a single PMU reading that systematically
     gives angle readings that are too large by 5 degrees, along with Gaussian
     random noise with standard deviation 0.0017. PMUs are at locations 4, 13,
     and 34. Huber scale parameter $\delta = 1.365 \cdot 0.0017$}
   \label{fig:badpmu}
\end{figure}

\begin{figure}[t]
   \center
   \input{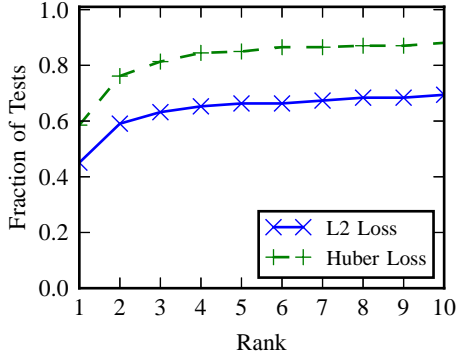}
   \caption{CDF for substation reconfigurations with Cauchy noise with a scale
     parameter of 0.001.}
   \label{fig:cauchy}
\end{figure}

In Figures \ref{fig:badpmu} and \ref{fig:cauchy} we show the usefulness of the
Huber loss in certain cases. In Figure \ref{fig:badpmu}, we run FLiER on the
line failures test (as in the noisy case of Figure \ref{fig:ieee57ranks}), but
with the PMU reading from bus 4 given a large post-event bias, always returning
an after-event angle reading that is 5 degrees too large. As shown in the
figure, this single bad reading can cause major problems with FLiER accuracy
using L2 loss. With Huber loss, however, that bad data is almost completely
ignored, resulting in accuracy essentially identical to that of the noisy case
in Figure \ref{fig:ieee57ranks}.

In Figure \ref{fig:cauchy}, we show the utility of the Huber loss in the
presence of heavy-tailed noise. Here we test substation reconfigurations with
noise given by a Cauchy distribution with scale parameter 0.001. As shown in the
figure, the use of the Huber loss significantly improves FLiER's performance
under this type of noise.

It is unclear if either of these situations are likely to arise in practice. In
the former case (with a bad PMU), it is likely that this bad data stream would
be identified in an earlier state estimation stage and already removed from the
fingerprint. In the latter case, the IEEE specification~\cite{6839218} requires that
the PMU noise profile not have heavy tails, so systematic heavy-tailed noise is
unlikely.

\subsection{Filter Effectiveness and Speed}

\begin{figure}[t]
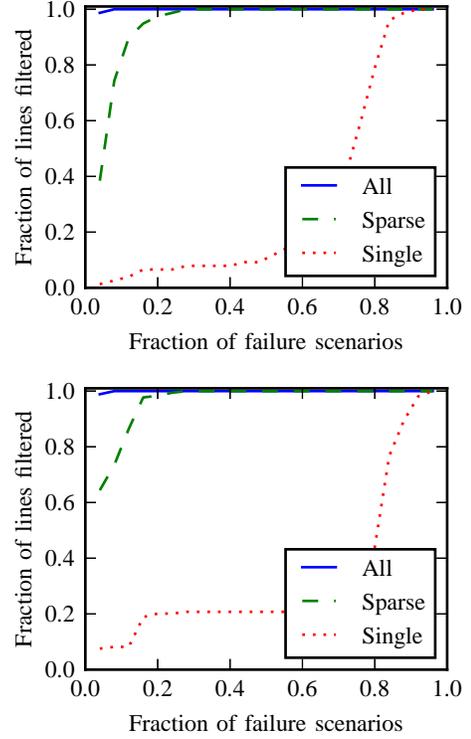

  \center
  \input{\MyPath data/ieee57lines_filter.pgf}
  \input{\MyPath data/ieee118lines_filter.pgf}
  \caption{Cumulative distribution function of fraction of lines for
    which $t_{ik}$ need not be computed when a line in the IEEE 57-bus
    (top) or 118-bus (bottom) network fails uniformly at random.}
  \label{fig:hist1}
\end{figure}

\begin{figure}[t]
  \center
  \input{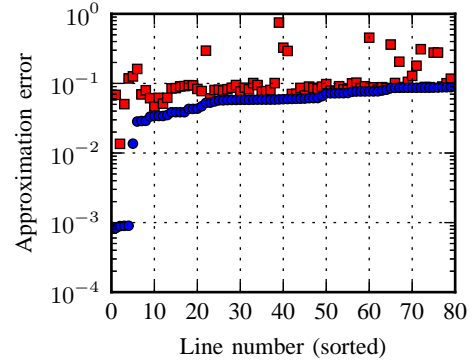}
  \caption{Example of effective filtering in Algorithm
    \ref{alg:fast_approx}. Each column represents a line checked. Blue
    dots are the lower bounds $\tau_{ik}$, while red squares are true
    scores $t_{ik}$. Columns are sorted by $\tau_{ik}$.  In this case,
    $t_{ik}$ only needs to be computed for eight lines.}
  \label{fig:filter_plot}
\end{figure}

The cost of FLiER depends strongly on the effectiveness of the
filtering procedure.  In Figure~\ref{fig:hist1}, we
show how often the filter saves us from computing fingerprint scores
in experiments on the IEEE 57-bus and 118-bus networks when checking
for line failures.  For each PMU deployment, we show the cumulative
distribution function of the fraction of lines for which fingerprint
scores need not be computed for each line failure.  The filter
performs well even for the sparse PMU deployments; we show a typical
case in Figure~\ref{fig:filter_plot}.

Finally, we demonstrate the importance of the filter by
running FLiER on the large Polish network~\cite{Zimmerman:2011:MATPOWER}
with 100 randomly placed PMUs.
Table~\ref{tab:polish_timing} shows
FLiER run times
with and without the filter
on ten randomly selected branches.
The code is unoptimized Python,
so these timings
do not indicate
of how fast FLiER
would run in a performance setting. However, they give a sense of the
speedup one expects from filtering.

\begin{table}
  \centering
  \begin{tabular}{| l | r |  r | r | r |}
    \hline
    Line & FLiER  (s) & Solution   &   \# t's          & FLiER n.f. (s) \\
         &                 & rank         &      computed &  \\
    \hline
    (1502, 917)   & 0.27 & 1 & 2  &  14.14  \\
    (1502, 1482) & 0.27  & 1 & 2  & 15.30  \\
    (557, 556)     & 0.27 & 1 & 4   &  14.75 \\
    (2346, 2341)  & 2.95 & 14 & 502 & 14.40  \\
    (909, 1155)    & 0.26  & 1 & 2  & 14.32  \\
    (644, 629)     & 0.29 & 1 & 7   & 14.59 \\
    (591, 737)    & 0.29 & 1 & 6  & 17.27  \\
    (559, 542)    & 0.32  & 1 & 13  & 16.59  \\
    (378, 336)    & 0.28 & 1 & 6  & 16.16 \\
    (101, 94)    & 0.26 & 1 & 2   & 15.29  \\
    \hline
  \end{tabular}
  \caption{FLiER run times for ten line failures with and without
    filtering.  About 3000 contingencies are considered.}
  \label{tab:polish_timing}
\end{table}

Note also that FLiER correctly identified the failed branch in 9 of 10
cases. In the one case in which it failed, on branch (2346, 2341),
$t_\ell$ for the correct answer was $6.25 \cdot 10^{-5}$; this
suggests
the failure had a negligible impact on the network.

We also performed this timing test for a randomly selected set of
substation reconfigurations, shown in
Table~\ref{tab:polish_substation_timing}.
Again, the results give a sense of the speedup one expects from filtering
in the substation reconfiguration case.

\begin{table}
   \centering
   \begin{tabular}{| l | r | r | r | r |}
      \hline
      Bus /           & FLiER (s) & Sol. rank /  & \# t's    & FLiER n.f. (s) \\
      Split nodes   &               & Sol. bus rank &  computed        & \\
      \hline
      86/1,2,3       &    4.54         &     1/1     & 2                   &    823.0 \\
      176/1,2        &    4.70          &     3/3    & 4                   &    836.9 \\
      539/7           &    8.11          &     1/1   & 38                   &    829.8 \\
      702/4,5        &    4.66          &     1/1    & 4                   &    820.1 \\
      754/2,3,4,5   &    4.97          &     1/1     & 7                   &    829.8 \\
      994/2,3,4,5   &    4.86          &     1/1     & 6                   &    835.8 \\
      1131/2          &    5.22          &     1/1     & 9                   &    850.3 \\
      1513/4,5,6     &    6.65          &     4/1     & 23                   &    862.3 \\
      1663/1,2,3,4  &    7.83          &     1/1     & 35                   &    928.1 \\
      2164/5          &    4.56          &     1/1     & 3                   &    875.3 \\
      \hline
   \end{tabular}
   \caption{FLiER run times for ten substation reconfigurations with
     and without filtering.  Nearly 7000 contingencies are considered.}
   \label{tab:polish_substation_timing}
\end{table}

\section{Conclusion and Future Work}

We have presented FLiER, a new algorithm to identify topology changes involving
load and generator trips, line outages, and substation reconfigurations using a
sparse deployment of PMUs.  Our method uses a linearization of the power flow
equations together with a novel subspace-based filtering approach to provide
fast diagnosis.  Unlike prior approaches based on DC approximation, our approach
takes advantage of a state estimate obtained shortly before the topology
changes, assuming that the network specifications remain unchanged or change in
a known way as a result of the failure.

Several extensions remain open for future work.  We hope to model
noise sensitivity of our computations, so that we can provide
approximate confidence intervals for fingerprint and filter scores; we
also believe it possible to diagnose when the linear approximation
will lead to incorrect diagnosis, and do more computation to deal just
with those cases.  In addition, we plan to extend our approach to
other events, such as single-phase line failures or
changes in line parameters due to overloading.

\bibliographystyle{IEEEtran}
\bibliography{power}

\end{document}